\documentclass[onecolumn,aps,prc,floatfix,showpacs,showkeys]{revtex4-1}
\usepackage{amsfonts}
\usepackage{amssymb}
\usepackage{amsmath}
\usepackage{amssymb}
\usepackage{graphicx}
\usepackage{dcolumn}
\usepackage{float}
\usepackage[T1]{fontenc}

\usepackage{hyperref}
\usepackage{appendix}
\usepackage{enumitem}

\begin{document}
\title{Least-squares fitting applied to nuclear mass formulas. \\  Resolution by the Gauss-Seidel method}
\author{B. Mohammed-Azizi} 
\email{azizi.benyoucef@univ-bechar.dz}
\affiliation{University of Bechar, Bechar, Algeria}
\affiliation{LPPPS Laboratory, Ecole Normale Supérieure, Vieux Kouba, Alger, Algeria }
\author{H. Mouloudj}
\affiliation{University of Chlef, Chlef, Algeria}
\affiliation{LPPPS Laboratory, Ecole Normale Supérieure, Vieux Kouba, Alger, Algeria }
\date{\today }
\begin{abstract}
 A numerical method optimizing the coefficients of the semi empirical mass formula or those of similar mass formulas is presented. The optimization is based on the least-squares adjustments method and leads to the resolution of a linear system which is solved  by iterations according to the Gauss-Seidel scheme. The steps of the algorithm are given in detail. In practice the method is very simple to implement and is able to treat large data in a very fast way. In fact, although this method has been illustrated here by specific examples, it can be applied without difficulty to any experimental or statistical data of the same type, i.e. those leading to linear system characterized by symmetric and positive-definite matrices.
\end{abstract}

\keywords{ Binding energy, Nuclear masses, Curve fitting}

\pacs{ 21.10.Dr, 02.60.Ed}
\maketitle

\section{Construction of the linear System of equations from a minimization
procedure}

\subsection{Principle of the method: Linear combination of independent functions}

The least squares method \cite{key-1,key-2} has historically attributed
to Gauss (1795). This method is applied in various scientific fields,
such as statistics, geodesy, economy, optimization, etc. .... The
present work proposes to apply it to the optimisation of the semi-empirical
\cite{key-3,key-4} mass formula or to formulas of the same type.
In this category, we can cite: (a) The FRDM (the best) and FRLDM models
with complete calculation of shell and pairing corrections by also considering
several kinds of nuclear deformations \cite{key-5}, (b) The ``Pomorski-Dudek
model'' \cite{key-6} with shell and pairing corrections, (c) The
``Royer model'' with shell and pairing corrections but without nuclear
deformation \cite{key-7}, (d) The well-known droplet model of Myers
based on the Thomas-Fermi approximation with and without shell correction
\cite{key-8}. Nuclear mass formulas are an important tool in the
evaluation of some ground-state properties, nuclear  reactions and in the
prediction of the neutron/proton drip lines. All these formulas
can be optimized in the same way by the method described by the following
procedure.

At the beginning, suppose that for each pair $(x_{i},y_{i})$ of independent
variables, we have to measure a correlated data $B_{exp}(x_{i},y_{i})$.
If we make $n$ measurements, we will have $i=1,2,......,n$ and in
practice we thus obtain a table with three columns of $n$ data, i.e. $x_{i}$, $y_{i}$, $B_{exp}(x_{i},y_{i})$. In the folllowing, for
the discrete indices of summation we adopt the general convenient ``rule''
that any discrete index such as for example $k$, will vary from $k=1$
(first data) to $k=kmax$ (last data). For example,  in the previous cited case the index
$"n"$ (last value of $i$) has been replaced by $imax$. In the same way, the index $j$
will vary from $1$ to $jmax$,  the index $k$
will vary from $1$ to $kmax$, , etc... \\ In this work,  we are looking for a mathematical
expression $B_{theo}(x_{i},y_{i})$ which can approximate ``as much
as possible'' the experimental data $B_{exp}(x_{i},y_{i})$. The
theoretical expression $B_{theo}(x_{i},y_{i})$ is defined by a suitable
linear combination of independent functions in which some coefficient have to be determined on the basis of the least squares method. This linear combination is of the following form:

\begin{equation}
B_{theo}(x_{i},y_{i})={\displaystyle \sum_{k=1}^{kmax}b_{k}F_{k}(x_{i},y_{i})+\sum_{j=1}^{jmax}c_{j}G_{j}(x_{i},y_{i})\ \ \ \ }i=1,imax\label{eq:1}
\end{equation}

In order to lighten the writing, we set: $B_{exp}(x_{i},y_{i})=B_{exp}(i)$,
$F_{k}(x_{i},y_{i})=F_{k}(i)$, $G_{j}(x_{i},y_{i})=G_{j}(i)$, $B_{theo}(x_{i},y_{i})=B_{theo}(i)$
.\\
 As mentioned before we will use the least squares method. Only the coefficients $b_{k}$ will be chosen
by variation (all other quantities are supposed to be known) so as to minimize the total sum of the squared deviations of
the theoretical values with respect to the experimental values.

\begin{equation}
E(b_{1},b_{2},...,b_{kmax})=\sum_{i=1}^{imax}\left(B_{exp}(i)-B_{theo}(i)\right)^{2}\label{eq:3}
\end{equation}

In Eq. (\ref{eq:1}), the coefficients $c_{j}$ will be maintained
constant. Thus, this linear combination contains a ``varied part''
and a ``constant part''. In the sums of Eq. (\ref{eq:1}) $kmax$
is the number of ``varied'' coefficients $b_{k}$ and $jmax$ is
the number of ``non varied'' coefficients $c_{j}$. With the "light" representation defined just above, $B_{theo}$ which appears in Eq. (\ref{eq:3}) and which is defined by equation (\ref{eq:1}), becomes:
\begin{equation}
B_{theo}(i)={\displaystyle \sum_{k=1}^{kmax}b_{k}F_{k}(i)+\sum_{j=1}^{jmax}c_{j}G_{j}(i),\ \ \ \ i=1,imax}\label{eq:2}
\end{equation}
Thus, according to the least squares principle, we must minimize the sum
given by Eq. (\ref{eq:3})
with respect to the quantities $b_{k}$. The minimum sought (we will
show later that this critical point is indeed a minimum) of this functional
corresponds to the system of equations defined by:

\begin{equation}
\frac{\partial E(b_{1},b_{2},...,b_{kmax})}{\partial b_{p}}=0\ \ \ \ \ \ \ \ p=1,2,.......,kmax\label{eq:4}
\end{equation}
In this expression, it is worth to repeat that all the quantities are known and fixed,
only the coefficients $b_{p}$ are unknown and are therefore determined
by variations (minimization). This leads to a system of $kmax$ equations with $kmax$
unknown quantities $b_{p}$ ( see Eq. (\ref{eq:4})). By definition
the function $E(b_{1},b_{2},...,b_{kmax})$ in Eq. (\ref{eq:3}) is
a linear combination of $b_{p}$, therefore it is continuous with
respect to these variables. We thus obtain by derivation and permutation
of summations, the following set of equations, each equation corresponds to a fixed
$p$ value:

\begin{equation}
-\sum_{i=1}^{imax}B_{exp}(i)F_{p}(i)+\sum_{k=1}^{kmax}b_{k}\sum_{i=1}^{imax}F_{k}(i)F_{p}(i)+\sum_{j=1}^{jmax}c_{j}\sum_{i=1}^{imax}G_{j}(i)F_{p}(i)=0\label{eq:5}
\end{equation}
It is convenient to set:

\begin{equation}
\sum_{i=1}^{imax}F_{k}(i)F_{p}(i)=\Phi_{kp},\label{eq:6}
\end{equation}

\begin{equation}
\sum_{i=1}^{imax}G_{j}(i)F_{p}(i)=\Gamma_{jp}\label{eq:7}
\end{equation}
\begin{equation}
\sum_{i=1}^{imax}B_{exp}(i)F_{p}(i)=\beta_{p}\label{eq:8}
\end{equation}
the previous set of equations become:

\begin{equation}
\sum_{k=1}^{kmax}b_{k}\Phi_{kp}=\beta_{p}-\sum_{j=1}^{jmax}c_{j}\Gamma_{jp}\ \ \ \ \ \ \ \ p=1,2,.......,kmax\label{eq:9}
\end{equation}
or in equivalent way, in matrix representation:

\begin{equation}
\left[\begin{array}{cccc}
\Phi_{11} & \Phi_{12} & .. & \Phi_{1q}\\
\Phi_{21} & \Phi_{22} & .. & \Phi_{2q}\\
.. & .. & .. & ..\\
\Phi_{q1} & \Phi_{q2} & .. & \Phi_{qq}
\end{array}\right]\left[\begin{array}{c}
b_{1}\\
b_{2}\\
..\\
b_{q}
\end{array}\right]=\left[\begin{array}{c}
\gamma_{1}\\
\gamma_{2}\\
..\\
\gamma_{q}
\end{array}\right]\label{eq:10-1}
\end{equation}
Here, $q=kmax$ and $\gamma_{p}=\beta_{p}-\sum_{j=1}^{jmax}c_{j}\Gamma_{jp}$.\\
 It is a linear system of $kmax$ equations with $kmax$ unknowns
(the $b_{k}$) whose the main determinant is that of the matrix $\Phi$
defined by Eq. (\ref{eq:6}). Thus the set of the unknowns $(b_{1},b_{2},...,b_{jmax})$
which minimize the total squared deviations given by Eq.(\ref{eq:3})
is the one which is the solution of the linear system given by  Eq. (\ref{eq:9}) or equivalently by Eq. (\ref{eq:10-1}).
Therefore, this minimization procedure amounts to solve a linear system
of equations .

\subsection{The system has always a solution}

It is clear from equation (\ref{eq:6}) that the matrix $\Phi$ is
real and symmetric. Let us define the rectangular matrix $F$ by its
matrix elements $F(i,k)=F_{k}(i)$, and let us calculate the product
$F^{T}F$ where $F^{T}$ is the transpose of the matrix $F$. So 
\begin{equation}
\left(F^{T}F\right)_{kp}=\sum_{j}F^{T}(k,j)F(j,p)=\sum_{j}F_{k}(j)F_{p}(j)=\Phi_{kp}\label{eq:10}
\end{equation}
therefore $F^{T}F=\Phi$. Thus, the matrix $\Phi$ can be set
in the form of a product of a matrix and its transpose. Moreover,  $F$ is constituted by $kmax$ linearly independent column-vectors.  Consequently
the matrix $\varPhi$ is real, symmetric and positive definite, it
is therefore invertible and the system given by Eq. (\ref{eq:9})
has always one solution and only one (i.e., the set $(b_{1},b_{2},...,b_{kmax})$)
which verify Eq. (\ref{eq:9})).

\subsection{Nature of the critical point}

In fact, the condition of Eq. (\ref{eq:4}) does not give necessarily
a minimum but corresponds generally to a critical point (minimum,
maximum or saddle point) . To show the nature of this critical point,
it is necessary to study the Hessian matrix $H_{pq}$ of the function
$E$ defined above by Eq. (\ref{eq:3}). The elements of the Hessian
matrix of $E$ are given by:

\begin{equation}
H_{pq}=\frac{\partial^{2}E(b_{1},b_{2},...,b_{kmax})}{\partial b_{p}\partial b_{q}}\label{eq:11}
\end{equation}
The first derivative is that given by the left member of Eq. (\ref{eq:5}).
A second derivative of $E$ gives:

\begin{equation}
H_{pq}=\frac{\partial^{2}E(b_{1},b_{2},...,b_{kmax})}{\partial b_{p}\partial b_{q}}=\Phi_{pq}\label{eq:12}
\end{equation}
The Hessian matrix $H$ is no more than the matrix $\Phi$. So the
Hessian matrix is symmetric, real and positive definite, therefore
the critical point (extremum) is always a minimum.

\section{Resolution by the Gauss-Seidel method }

\subsection{Gauss-Seidel vs some other methods}

Compared to direct solution methods, iterative methods become indispensable
when the size of the system becomes large. Indeed, direct methods
require a number of floating point operations of the order of $n^{3}$
($n$ is the size of the system, i.e. the number of equations) tends
to infinity which makes them slow for large values of $n$. In this
case direct methods must be avoided and methods such as Jacobi's,
Gauss-Seidel and conjugate gradient must be preferred. In practice,
the implementation of a numerical method is always a compromise between
several advantages and inconvenients. For simplicity of programming,
the Jacobi and Gauss-Seidel methods prevail over the conjugate gradient
method. Compared to the method of Jacobi, that of Gauss-Seidel is
not suitable for parallelization. But this advantage is of little
interest here because the iterations are based on a single short formula and also because the Gauss-Seidel method converges
faster than that of Jacobi. Finally, Gauss-Seidel method was preferred. Because, the
matrix $\Phi$ of the system (\ref{eq:9}) is symmetrical and definite-positive,
the Gauss-Seidel method always converges whatever the starting point
is. The convergence is thus guaranteed.

\subsection{Algorithm of the iterative procedure}

In Eq. (\ref{eq:9}) we can isolate the term $k=p$ expressing thus
the coefficient $b_{p}$ as a linear combination of the remaining
terms:

\begin{equation}
b_{p}=\frac{\beta_{p}-{\displaystyle \sum_{k\neq p}^{kmax}}b_{k}\Phi_{kp}-{\displaystyle \sum_{j=1}^{jmax}}c_{j}\Gamma_{jp}}{\Phi_{pp}}\ \ \ \ \ \ \ \ p=1,2,.......,kmax\label{eq:13}
\end{equation}
It should be noted that each unknown $b_{p}$ appears only in the
left-hand side of the equation.It is thus only expressed by the remaining
set of unknowns. The denominator is by definition:

\begin{equation}
\Phi_{pp}=\sum_{i=1}^{imax}\left(F_{p}(i)\right)^{2}\label{eq:14}
\end{equation}
It is then strictly positive (the only exception would be that for
fixed $p$, the $F_{p}(i)$ are all zero, which would be equivalent
to taking a null function, such a choice would be absurd). So the
denominator does not present a problem.\\
 The Gauss-Seidel method consists in successive calculations, 
$b_{1}$, $b_{2}$, $b_{3}$,.... In each iteration the new value replaces the old value of the previous according to the following steps:\\
 The first ``round'' corresponds to the $kmax$ first iterations.
The successive iterations (determinations of $b_{k}$) follow the
scheme:

\[
\begin{array}{c}
(b_{1}^{0},b_{2}^{0},...,b_{kmax}^{0})\rightarrow b_{1}^{1}\\
(b_{1}^{1},b_{2}^{0},...,b_{kmax}^{0})\rightarrow b_{2}^{1}\\
(b_{1}^{1},b_{2}^{1},...,b_{kmax}^{0})\rightarrow b_{3}^{1}\\
...\\
(b_{1}^{1},b_{2}^{1},...,b_{kmax}^{1})\rightarrow b_{1}^{2}
\end{array}
\]
here $b_{p}^{k}$ corresponds to the iteration number $k$ of the
pth coefficient of the unknown set. In fact one round corresponds
simply to one iteration for all the unknowns of the set $(b_{1},b_{2},...,b_{kmax})$.
\\
 The second \textquotedbl round\textquotedbl{} (which is also a set
of $kmax$ iterations) is defined by similar iterations, starting
from the last iteration of the first \textquotedbl round\textquotedbl .

\[
\begin{array}{c}
(b_{1}^{1},b_{2}^{1},...,b_{kmax}^{1})\rightarrow b_{1}^{2}\\
(b_{1}^{2},b_{2}^{1},...,b_{kmax}^{1})\rightarrow b_{2}^{2}\\
(b_{1}^{2},b_{2}^{2},...,b_{kmax}^{1})\rightarrow b_{3}^{2}\\
...\\
(b_{1}^{2},b_{2}^{2},...,b_{kmax}^{2})\rightarrow b_{1}^{3}
\end{array}
\]
etc, ..... This process continue until convergence, i.e. until a stable set $(b_{1}^{n},b_{2}^{n},...,b_{kmax}^{n})$
after a sufficient number $n$ of ``rounds''.\\
 Thus each round is constituted by $kmax$ iterations, in each iteration
one of the coefficient $b_{p}$ is determined by Eq. (\ref{eq:13})
and the old value is ``immediately'' replaced by the new one.\\
 Obviously, at the starting of the procedure, an initial set $(b_{1}^{0},b_{2}^{0},...,b_{kmax}^{0})$
must be chosen. Moreover, we know from the beginning of the present
section that the iterative procedure always converges, whatever the
values of the initial set. However, in order not to slow down the
process by an awkward choice at the start of the iterations, it is
convenient to start from a ``suitable'' set chosen for example on
the basis of physical considerations.

In practice, the most important point in this method, is to store
once and for all the matrices $\Phi{}_{kp}$, $\Gamma_{jp}$ and the
vector $\beta_{p}$ at the start of the program before the beginning
of the iterations. It is also possible to use the symmetry of $\Phi_{kp}$
to reduce greatly the calculations. Moreover it should be noted that
the matrix $\Gamma_{jp}$ (see Eq. (\ref{eq:7})) which corresponds
to the coefficients which are not ``varied'' is in general much
smaller than the matrix $\Phi_{kp}$ because in order to obtain an efficient optimization, it is more advantageous
to vary the maximum number of parameters. Indeed, due to physical
constraints only some parameters are maintained constant. The present
considerations leads to simplify greatly the programming and the efficiency
of the code.

\section{Illustrations of this method }

In fact this algorithm applies to any linear system, but here the
main motivation was to implement it to the optimization of the coefficients
of mass formulas of nuclear physics.\\
 In this numerical study we will consider two cases: The simple case
of the semi empirical mass formula and a more elaborate version of
the liquid drop mass model containing a larger number of corrective
terms.

\subsection{The Bethe-Weizsaker formula}

\subsubsection{Steps of the algorithm for a  simple case}

The semi empirical mass formula or Bethe-Weizsaker formula \cite{key-3,key-4}
is well known. It approaches theoretically the binding energy of the
atomic nucleus only as a function of the number of $N$ neutrons and
$Z$ protons. The theoretical mass of the nuclei can be deduced straightforwardly
from this binding energy. The standard Bethe-Weiszaker formula gives
the theoretical expression of the binding energy $B_{theo}(N,Z)$
of the atomic nucleus as a function of a five contribution:

\begin{equation}
B_{theo}(A,Z)\approx a_{V}A-a_{S}A^{2/3}-a_{C}Z^{2}/A^{1/3}-a_{a}(A-2Z)^{2}/A+\delta(A,Z)\label{eq:15}
\end{equation}
With $A=N+Z$. The first four terms are respectively the well known
volume, surface, Coulomb and asymmetry energies. The fifth is the
pairing energy term:

\begin{equation}
\delta(A,Z)=\varepsilon12.0/A^{1/2}\label{eq:16}
\end{equation}
where $\epsilon=0$ if $A$ odd , $\varepsilon=+1$ if $N$ and $Z$
even, $\varepsilon=-1$ if $N$ and $Z$ odd\\
 Here the procedure was done in FORTRAN 90 language but due to the
simplicity of the algorithm described here it is obviously possible
to use other language codes. The goal of the method is to solve Eq.
(\ref{eq:13}) in order to find the four optimized constants $\left(a_{V},a_{S},a_{C},a_{a}\right).$
In this typical example, the steps of the algorithm are illustrated
as follows: 
\begin{itemize}
\item Reading and storing the values of numbers of neutrons and protons $(N_{i},Z_{i})$ and their experimental
binding energy from an input data file. 
\item Construction of the four functions $F_{k}(x_{i},y_{i})$ of Eq.(\ref{eq:6}-\ref{eq:8})
corresponding to the varied coefficient ($k=1,4$):\\
 Remembering that for each nucleus ``number $i$'' represents a
couple $(N_{i},Z_{i})$ and setting $A_{i}=N_{i}+Z_{i}$., we will
have here; 
\end{itemize}
\begin{equation}
\begin{array}{c}
F_{1}(i)=F_{1}(N_{i},Z_{i})=A_{i}\\
F_{2}(i)=F_{2}(N_{i},Z_{i})=-A_{i}^{2/3}\\
F_{3}(i)=F_{3}(N_{i},Z_{i})=-Z_{i}^{2}/A_{i}^{1/3}\\
F_{4}(i)=F_{4}(N_{i},Z_{i})=-(A_{i}-2Z_{i})^{2}/A_{i}\\
\\
\end{array}\label{eq:17}
\end{equation}

\begin{itemize}
\item Construction of the function $G_{1}(i)=G_{1}(x_{i},y_{i})=\varepsilon a_{p}/A_{i}^{1/2}$
(Eq. (\ref{eq:7})), there is here only one function $G$. We set
$c_{1}=1$ since this coefficient is not varied. 
\item Before the iterations, the following matrices must be calculated and
stored in the memory of the computer: 
\end{itemize}
\begin{equation}
\begin{array}{c}
\Phi_{kp}=\sum_{i=1}^{imax}F_{k}(i)F_{p}(i)\ \ \ \ \ \ \ \ k,p=1,4\\
\Gamma_{jp}=\sum_{i=1}^{imax}F_{j}(i)G_{p}(i)\ \ \ \ \ \ \ \ j=1,4\ \ \ \ \ \ \ \ p=1\\
\beta_{p}=\sum_{i=1}^{imax}B_{exp}(i)F_{p}(i)\ \ \ \ \ \ \ \ p=1,4
\end{array}\label{eq:18}
\end{equation}
$B_{exp}(N_{i},Z_{i})=B_{exp}(i)$ is the experimental binding energy
of the ``i-th'' nucleus, the summation is done over $i$ up to the
number of nuclei which is equal to $imax$. 
\begin{itemize}
\item Choosing a starting set $(b_{1}^{0},b_{2}^{0},.b_{3}^{0},.b_{4}^{0})=(a_{V}^{0},a_{S}^{0},a_{C}^{0},a_{a}^{0})$ 
\item Use Eq. (\ref{eq:13}) to determine the unknowns by an iterating sequence
as follows: 
\end{itemize}
\[
\begin{array}{c}
(a_{V}^{0},a_{S}^{0},a_{C}^{0},a_{a}^{0})\rightarrow a_{V}^{1}\\
(a_{V}^{1},a_{S}^{0},a_{C}^{0},a_{a}^{0})\rightarrow a_{S}^{1}\\
(a_{V}^{1},a_{S}^{1},a_{C}^{0},a_{a}^{0})\rightarrow a_{C}^{1}\\
(a_{V}^{1},a_{S}^{1},a_{C}^{1},a_{a}^{0})\rightarrow a_{a}^{1}\\
(a_{V}^{1},a_{S}^{1},a_{C}^{1},a_{a}^{1})\rightarrow a_{V}^{2}\\
(a_{V}^{2},a_{S}^{1},a_{C}^{1},a_{a}^{1})\rightarrow a_{S}^{2}\\
....
\end{array}
\]

etc.... A new set (four iterations in this case) is calculated in
one round as explained above. 
\begin{itemize}
\item Stop the process as soon as successive iterations does not change
the results anymore (up to a desired reasonable fixed precision) 
\item Evaluating root mean square deviation from experimental 
\end{itemize}

\subsubsection{Results: Speed, convergence rate and efficiency of the method}

The experimental binding energy of each nuclei is extracted from the
recent updated Atomic Mass Evaluation, i.e. AME 2020 table, published
in \cite{key-9,key-10} . Non-experimental (i.e. estimated) values
are not taken into account considering only nuclei with $N\geq8$
and $Z\geq8$. In this case the number of nuclei ($imax$) for the
least square fit is equal to 2452. The atomic masses are given
with a maximum uncertainty of $\pm0.0009u$ (unified mass unit). \\
 As usual, the root mean square deviation is defined by:

\begin{equation}
\sigma=\dfrac{\left(\sum_{i=1}^{imax}\left(B_{exp}(i)-B_{theo}(i)\right)^{2}\right)^{1/2}}{imax}\label{eq:19}
\end{equation}
where $i=(N_{i},Z_{i})$ and $B_{theo}$ is given by Eq. (\ref{eq:15}).
The calculations are made on a desktop P.C. computer with Intel i3
4160 processor, and a memory of 16 Gbytes (DDR3 type).

To study convergence, the number of total iterations was set at 80,000,000
(80 million of iterations). However, in spite of the number of iterations,
the execution time is only about 21 seconds for this case where only
four quantities where varied. The values of the coefficients are recorded
for a certain number of iterations (see table \ref{tab: table1} ).
For this small system, one notes a stability of the results for the
four coefficients starting from 100000 iterations. Here, stability
is defined arbitrarily by requiring that the 8 digits after the decimal
point no longer change . With these stabilized values we obtain a
root mean square deviation of 3.06 MeV in the binding energy. The
``optimized'' formula is thus:

\begin{equation}
\begin{split}B_{theo}(A,Z)\approx15.5461A-16.9598A^{2/3}\\
-0.7036Z^{2}/A^{1/3}-23.0253(A-2Z)^{2}/A+\delta(A,Z)
\end{split}
\label{eq:20}
\end{equation}
It is well known that the basic standard empirical formula with five
terms of which four coefficients are optimized does not allow great
precision. A refinement is possible with the increase of the number
of corrective terms and the introduction of the shell and pairing
corrections. This greatly improves the precision (see below the liquid
drop formula).

\begin{table}
\begin{tabular}{|c|c|c|c|c|}
\hline 
Iterations  & $a_{V}$  & $a_{S}$  & $a_{C}$  & $a_{a}$\tabularnewline
\hline 
\hline 
100  & 5.70152423  & 13.20519299  & 0.02379798  & 0.61415300\tabularnewline
\hline 
500  & 7.26660331  & 8.48293937  & 0.13373169  & 4.25226639\tabularnewline
\hline 
1000  & 8.86615851  & 3.56756982  & 0.24383570  & 7.87910579\tabularnewline
\hline 
5000  & 14.34684644  & 13.27436635  & 0.62109416  & 20.30604482\tabularnewline
\hline 
10000  & 15.40599727  & 16.52909474  & 0.69399990  & 22.70756867\tabularnewline
\hline 
50000  & 15.54615806  & 16.95980328  & 0.70364775  & 23.02536994\tabularnewline
\hline 
100000  & 15.54615806  & 16.95980329  & 0.70364775  & 23.02536995\tabularnewline
\hline 
500000  & 15.54615806  & 16.95980329  & 0.70364775  & 23.02536995\tabularnewline
\hline 
1000000  & 15.54615806  & 16.95980329  & 0.70364775  & 23.02536995\tabularnewline
\hline 
2000000  & 15.54615806  & 16.95980329  & 0.70364775  & 23.02536995\tabularnewline
\hline 
3000000  & 15.54615806  & 16.95980329  & 0.70364775  & 23.02536995\tabularnewline
\hline 
4000000  & 15.54615806  & 16.95980329  & 0.70364775  & 23.02536995\tabularnewline
\hline 
5000000  & 15.54615806  & 16.95980329  & 0.70364775  & 23.02536995\tabularnewline
\hline 
10000000  & 15.54615806  & 16.95980329  & 0.70364775  & 23.02536995\tabularnewline
\hline 
15000000  & 15.54615806  & 16.95980329  & 0.70364775  & 23.02536995\tabularnewline
\hline 
20000000  & 15.54615806  & 16.95980329  & 0.70364775  & 23.02536995\tabularnewline
\hline 
60000000  & 15.54615806  & 16.95980329  & 0.70364775  & 23.02536995\tabularnewline
\hline 
80000000  & 15.54615806  & 16.95980329  & 0.70364775  & 23.02536995\tabularnewline
\hline 
\end{tabular}

\caption{\label{tab: table1}Iterations and convergence of the four coefficients
of the Bethe-Weizsaker formula }
\end{table}

\subsection{The liquid drop mass model}

The liquid drop mass formula gives better results. In this respect
it is worth to mention that there are several versions of this model
. One of such improved formula is given by the following equation
which is very close to the ones \cite{key-5,key-6,key-7} (and several
references quoted therein):

\begin{equation}
\begin{split}B_{theo}(A,Z)\approx a_{V}(1.0-k_{V}I^{2}-k_{V}^{'}I^{4})A\\
-a_{S}(1.0-k_{S}I^{2}-k_{S}^{'}I^{4})A^{2/3}B_{S}(\beta)\\
-a_{curv}(1.0-k_{curv}I^{2}-k_{curv}^{'}I^{4})A^{1/3}B_{curv}(\beta)\\
-a_{C}(Z^{2}/A^{1/3})B_{coul}(\beta)+f_{p}(Z^{2}/A)B_{coul}(\beta)\\
+a_{exch}Z^{2}A^{1/3}+W\left|I\right|+a_{0}A^{0}+\varepsilon_{p}+E_{cong}\\
\\
\end{split}
\label{eq:21}
\end{equation}

The different versions of the liquid drop mass models are justified
through various theories and approximation (but this is beyond the
scope of this paper). The volume energy is given by the term containing
the constant $a_{V}$ and then we have the surface energy (with constant $a_{S})$.
These two contributiosn contain a part of the asymmetry energy defined
in the Bethe-Weizsaker formula seen before. Here, the relative neutron
excess is $I=(N-Z)/A$. The three following contributions correspond
to the curvature contribution ($a_{curv})$, the Coulomb energy ($a_{c})$
and the the diffuseness correction term $(f_{p})$ of the Coulomb
energy. The other three following terms, are respectively the exchange Coulomb
energy ($a_{exch})$ , the Wigner correction $W$, and the $A^{0}$
correction. At last, the two last contributions are (i) the smooth
pairing energy defined as before and (ii) the congruence energy given
by $E_{cong}=10exp(-4.2I^{2})$ . The surface, curvature and Coulomb
energy are taken deformation dependent through the quantities $B_{S}(\beta)$,
$B_{curv}(\beta)$ and $B_{coul}(\beta)$ where $\beta$ is the quadrupole
deformation as defined in Ref. \cite{key-5}. All these quantities can be
found in the Ref. \cite{key-5,key-6,key-7}. Anyway this version of
the liquid drop mass formula contains 16 contributions in which 14
are varied (the $b_{k}$). The ``family'' of models in which shell
and pairing corrections are taken into account belongs to the so-called
Microscopic-Macroscopic (MicMac) Model.

Whatever the number of elements involved in this formula, the steps
of the algorithm are strictly the same as the one seen above. We will
avoid repeating the operations described above. We only will confine
ourselves to see the performances in time (speed) and in quality (precision)
for this case.

\begin{table}
\begin{tabular}{|c|c|c|c|}
\hline 
Iterations  & $a_{V}$  & $a_{S}$  & $a_{C}$\tabularnewline
\hline 
\hline 
100  & -80.04729872  & -34.25277039  & -5.13507732\tabularnewline
\hline 
500  & -16.05087356  & 15.24624851  & -7.86998599\tabularnewline
\hline 
1000  & -7.37605620  & 16.42207924  & -8.84936291\tabularnewline
\hline 
5000  & 25.81407742  & 43.71052549  & -11.51082339\tabularnewline
\hline 
10000  & 46.71666775  & 66.73928076  & -12.17695777\tabularnewline
\hline 
50000  & 56.60653388  & 80.29100831  & -9.72429294\tabularnewline
\hline 
100000  & 42.11538222  & 86.75529165  & -7.17358506\tabularnewline
\hline 
500000  & 11.31733037  & 51.58243500  & -1.01117089\tabularnewline
\hline 
1000000  & 12.56436077  & 12.17115008  & -0.60429110\tabularnewline
\hline 
2000000  & 15.34111642  & -15.00144952  & -0.68255779\tabularnewline
\hline 
3000000  & 16.11176323  & -22.18537006  & -0.71035146\tabularnewline
\hline 
4000000  & 16.32071582  & -24.13250365  & -0.71790041\tabularnewline
\hline 
5000000  & 16.37737312  & -24.66046570  & -0.71994733\tabularnewline
\hline 
10000000  & 16.39842008  & -24.85659217  & -0.72070772\tabularnewline
\hline 
15000000  & 16.39845093  & -24.85687963  & -0.72070884\tabularnewline
\hline 
20000000  & 16.39845098  & -24.85688005  & -0.72070884\tabularnewline
\hline 
40000000  & 16.39845098  & -24.85688005  & -0.72070884\tabularnewline
\hline 
80000000  & 16.39845098  & -24.85688005  & -0.72070884\tabularnewline
\hline 
\end{tabular}

\caption{\label{tab:table 2}Iterations and convergence of three coefficient
of the liquid drop mass formula}
\end{table}

This time, the calculations take approximately 91 seconds which is
a good time if we consider that 80 millions of iterations are performed
with a set of 14 unknowns. Table (\ref{tab:table 2}) shows the convergence
of the three quantities $(a_{V},a_{S},a_{C})$ among the
fourteens unknown of Eq. (\ref{eq:21}). It is to be noted that ``8-digit stability
after the decimal point'' is reached after 20,000,000 (20 million)  iterations. The root mean square deviation is then about $1.81$
MeV. Compared to the semi empirical, the precision is greatly improved
($\sigma\approx1.81$ MeV against $\sigma\approx3.06$ MeV). The resulting
formula is given by Eq. (\ref{eq:22})

\begin{equation}
\begin{split}B_{theo}(A,Z)\approx16.3984(1.0-9.65912I^{2}-281.1899I^{4})A\\
-24.8568(1.0+125.7999I^{2}-2318.6417I^{4})A^{2/3}B_{S}(\beta)\\
+17.8930(1.0+358.6967I^{2}-4640.8412I^{4})A^{1/3}B_{curv}(\beta)\\
-0.720708(Z^{2}/A^{1/3})B_{coul}(\beta)+1.54147(Z^{2}/A)\\
-0.000558043Z^{2}A^{1/3}-19.6944\left|I\right|-24.4926A^{0}+\varepsilon_{p}+E_{cong}\\
\\
\end{split}
\label{eq:22}
\end{equation}

If we take into account the shell and pairing corrections (evaluated in a separate theoretical calculations), the root mean square deviation
is in a dramatic way reduced to $\sigma\approx0.864$ MeV. This explains
why shell and pairing corrections are absolutely necessary in this kind of formulas.

It is worth to emphasize that direct comparison between various type
of mass formulas proposed in the litterature has only a relative meaning
because very often the ``basis'', i.e. the set and the number of
nuclei is not the same. There are other factors involved, such as
the fact that the microscopic corrections are model dependent. Moreover,
in some serious calculations, the root mean square deviation is weighted
by some measurement error, etc...

\section{Conclusion}

The goal of the present work consisted essentially in determining
the best linear combination (i.e. the best coefficients,) of suitable
functions intended to approximate the experimental binding energy
of atomic nuclei. A very simple algorithm based on the Gauss-Seidel
method is described in detail. The method is characterized by the
maximum of simplicity in the procedure. This simplicity is the guarantee
of efficiency and speed on a practical level.

\end{document}